\documentclass{jfm}

\usepackage{hyperref}

\usepackage{graphicx}
\usepackage{epstopdf, epsfig}

\usepackage{amsmath}
\usepackage{amsfonts}
\usepackage{amssymb}
\usepackage{wasysym}
\usepackage{float}
\usepackage{color}

\shorttitle{Explosive ripple instability due to incipient wave breaking}
\shortauthor{A.A. Mailybaev and A. Nachbin}

\title{Explosive ripple instability due to incipient wave breaking}

\author{Alexei A. Mailybaev\corresp{\email{alexei@impa.br}} \and Andr\'e Nachbin}

\affiliation{Instituto Nacional de Matem\'atica Pura e Aplicada -- IMPA, Rio de Janeiro, Brazil}

\begin{document}

\maketitle

\begin{abstract}
Considering two-dimensional potential ideal flow with free surface and finite depth, we study the dynamics of small-amplitude and short-wavelength wavetrains propagating on the background of a steepening nonlinear wave. This can be seen as a model for small ripples developing on slopes of breaking waves in the surf zone. Using the concept of wave action as an adiabatic invariant, we derive an explicit asymptotic expression for the change of ripple steepness. Through this expression, nonlinear effects are described using the intrinsic frequency and intrinsic gravity along Lagrangian (material) trajectories on a free surface. We show that strong compression near the tip on the wave leads to an explosive (super-exponential) ripple instability. This instability may play important role for understanding fragmentation and whitecapping at the surface of breaking waves. Analytical results are confirmed by numerical simulations using a potential theory model.
\end{abstract}


\section{Introduction}
\label{sec1}

Water waves propagating on a moving  background, as for example a mean-flow or a current, is a problem of great interest from both the theoretical as well as the applications' viewpoints. 
Two recent books are exactly within these approaches and contain a great deal of references for the interested reader. The book  by~\cite{ConstantinSIAM}, {\it Nonlinear Water Waves with Applications to Wave-Current Interactions and Tsunamis}, 
presents the analysis recently developed for proving, for example, the existence of nonlinear periodic
waves in the presence of an underlying flow of constant vorticity, using global bifurcation and degree theory.  
Also of interest is the analysis of particle trajectories beneath these nonlinear waves, where closed orbits, stagnation points and critical layers can exist~\citep{constantin2016global}; see also~\citep{nachbin2014boundary} for the numerical studies.
The second book by~\cite{Buhler}, {\it Waves and Mean Flows}, approaches this topic from another perspective. It combines asymptotic methods, using for example averaging and ray tracing techniques, 
within either the Eulerian or Lagrangian description. Methods of non-canonical Hamiltonian mechanics are discussed regarding topics such as dispersive ray tracing in a moving frame, adiabatic invariance and wave action conservation. 

The concept of wave action in fluid dynamics, which we use as the main analytical tool in our work, goes back to the classical work by~\cite{bretherton1968wavetrains}, who presented a general formulation for wave-action conservation given a nonuniform underlying flow. 
Their approximate theory, analogous to the WKB approximation, considers a slowly varying wavetrain, of small amplitude, propagating in a non-homogenous moving media. 
The wave action is defined as the ratio $E/\Omega$ of the wave energy density $E$ by the 
intrinsic frequency $\Omega$, computed along rays. The intrinsic frequency 
(or relative frequency) 
 is the frequency measured in the moving reference frame of the local mean flow.
 Accounting for a Doppler shift, it is expressed as
\begin{equation}
\Omega = \omega - U k, 
\label{Doppler}
\end{equation}
where $\omega$ is the frequency, $k$ is the wavenumber and $U$ is the medium's local speed relative to the observer. \cite{lighthill2001waves} calls attention to the fact that
this scenario can modify the energetics of the wave propagation: wave energy increases (at the expense  of the mean flow) whenever the rays move into regions of greater
$\Omega$. 

In the present work we use the same wave action principles but, 
 in connection with nonlinear effects, our route to obtain the intrinsic quantities in the wave action is different.
In our formulation the background flow is taken to be that on the surface of a steepening nonlinear wave, while the propagating wavetrain refers to ripples -- small-amplitude and short-wavelength surface perturbations.
As will be shown, in order to capture nonlinear effects acting on the ripple, one needs an extra quantity $g_\ast$ representing the local intrinsic gravity.
In this fashion we are able to accurately account for a strongly accelerated background flow which,
through strong compression, leads to an explosive (super-exponential) ripple instability at the
expense of this mean flow. The intrinsic gravity is obtained 
from the (nonlinear) momentum equation written in Lagrangian form. Once $g_\ast$ is obtained
the respective intrinsic frequency $\Omega$ is readily available from the dispersion relation and we can compute the 
wave action $E/\Omega$.
Finally, using wave action conservation, we  present an analytic expression for the evolution of ripple's steepness. This expression shows that the explosive instability is promoted when the modulated wave moves into regions where the
surface particle trajectories display a strong
compression pattern. This compression induces an increase in the ripple
wavenumber and amplitude. 
Our simple expression captures this growth very accurately when compared to 
the numerical simulations using a potential theory model.
  
Note that a rigorous theory
for the problem here presented is not yet available and is desirable for a better understanding, for example, of the parameter range where the instability takes place, as well as the possible applications in the ocean. 
An important (potential) application is discussed next, which considers the ripple instability as an integral part of the complex multi-scale process of wave surface fragmentation and whitecapping~\citep{villermaux2007fragmentation,dyachenko2016whitecapping}. 
Such application requires further investigations, which at this stage are beyond the scope of the present work.

This paper is organized as follows. In section 2 we describe the features of marker dynamics on the surface of the steeping wave.
Section 3 presents the wave action formulation and our new simple formula for the ripple instability, expressing the
explosive (super-exponential) growth of the ripple's steepness. Numerical results are presented in section 4. A remarkable 
agreement is observed  between our formula and simulations with the nonlinear potential theory equations.
The conclusions are given in section 5, and the Appendix contains details of the numerical method.

\section{Lagrangian dynamical features at the onset of wave breaking}\label{sec2}

Our goal is to study dynamical features for  a  small ripple riding on the front face of a nonlinear breaking wave. First we will describe, through numerical simulations, Lagrangian properties (of material points) on the surface of a breaking wave. 
Then we consider the added ripple.

The phenomenon of gravity wave  breaking  is described by the Euler equations
	\begin{equation}
	\mathbf{v}_t+\mathbf{v}\cdot\nabla\mathbf{v} = -\nabla p/\rho+\mathbf{g},
	\qquad
	\mathrm{div}\,\mathbf{v} = 0,
	\label{eq1}
	\end{equation}
where the effects of viscosity and surface tension are neglected; comments on capillary effects will be made later. In the two-dimensional formulation, $(x,y)$ are the horizontal and vertical coordinates, $\mathbf{v} = (u,v)$ is the fluid velocity satisfying the incompressibility condition, $\rho$ is the constant density and $\mathbf{g}$ is the vector of gravitational acceleration. We will consider the potential flow over a flat rigid bottom, which is bounded from above by a free surface $y = F(x,t)$. A dimensionless formulation of the potential theory equations \citep{landau1987fluid,whitham2011linear} will be considered with a unit water depth, density and gravity acceleration. For numerical convenience, we assume that the flow is periodic in the horizontal direction with period $2\pi$, and set the rigid bottom coordinate at $y = -1$. 

We write the (kinematic and dynamic) boundary conditions at the free surface as
	\begin{equation}
	y = F(x,t):\quad v = F_t+F_x u,\quad p = P_{atm},
	\label{eq1bcU}
	\end{equation}
where $P_{atm}$ is a constant atmospheric pressure, and the subscripts $t$, $x$ and $y$ are used in this section to denote partial derivatives.
At the bottom, we have 
	\begin{equation}
	y = -1:\quad v = 0.
	\label{eq1bcB}
	\end{equation}

For potential incompressible flow, we can introduce the complex potential $\Phi = \varphi+i\psi$, which  is a holomorphic function of $z = x+iy$ in the fluid domain. The real potential function $\varphi(x,y,t)$ and the stream function $\psi(x,y,t)$ are related to the fluid velocities as 
	\begin{equation}
	u = \varphi_x = \psi_y,\qquad
	v = \varphi_y = -\psi_x. 
	\label{eq1pot}
	\end{equation}
Expressing velocities from (\ref{eq1pot}) and the pressure from the Bernoulli equation, the boundary conditions (\ref{eq1bcU})--(\ref{eq1bcB}) can be written as~\citep{landau1987fluid,whitham2011linear} 
	\begin{equation}
	y = F(x,t):\quad \varphi_y = F_t+F_x\varphi_x,\quad 
	\varphi_t+\frac{1}{2}\left(\varphi_x^2+\varphi_y^2\right)+gy = 0,
	\label{eq1bcU2}
	\end{equation}
	\begin{equation}
	y = -1:\quad \varphi_y = 0.
	\label{eq1bcB2}
	\end{equation}

Let $z(\zeta,t)$ with 
$\zeta = \xi+i\eta$ be a conformal mapping from a horizontal strip $-K \le \eta \le 0$ onto the fluid domain at time $t$. Such a mapping provides the free surface parametrization as $x+iy = z(\xi,t)$, with a real coordinate  $\xi$. 
Note that this mapping does not require that the free surface equations can be resolved with respect to the vertical coordinate, $y = F(x,t)$, i.e., it can be used when the free boundary has overhanging sections. 
With the method of complex analysis \citep{dyachenko1996nonlinear,zakharov2002new,ribeiro2017flow} one can describe the flow in the whole fluid domain in terms of real functions defined at the free-surface; see the Appendix \S\ref{secA1}. In this description, 
the equations of motion reduce to nonlocal differential equations in one spatial dimension $\xi$ and time $t$. 
This setting is very convenient for simulating numerically the potential theory equations
taking advantage of properties of harmonic functions in a strip.

The following numerical results illustrate the wave overturning;
details of the numerical method are presented in the Appendix \S\ref{secA2}.
Figure \ref{fig1} shows a familiar overturning wave profile above a flat rigid bottom at $y = -1$. Of  particular interest at this stage, we demonstrate the surface compression near the tip of a breaker displayed by numerical markers. 
We have chosen the initial profile $y = 0.35\cos x$ and the velocity potential at the surface $\varphi = ({0.35}/{\sqrt{\tanh 1}}) \sin x$, motivated by the linear theory. We will use this specific initial profile for all numerical simulations throughout the  paper. We also performed simulations for different initial conditions (not reported here) and observed qualitatively the same results for all aspects discussed in this work. 

\begin{figure}
\centering
\includegraphics[width=0.8\textwidth]{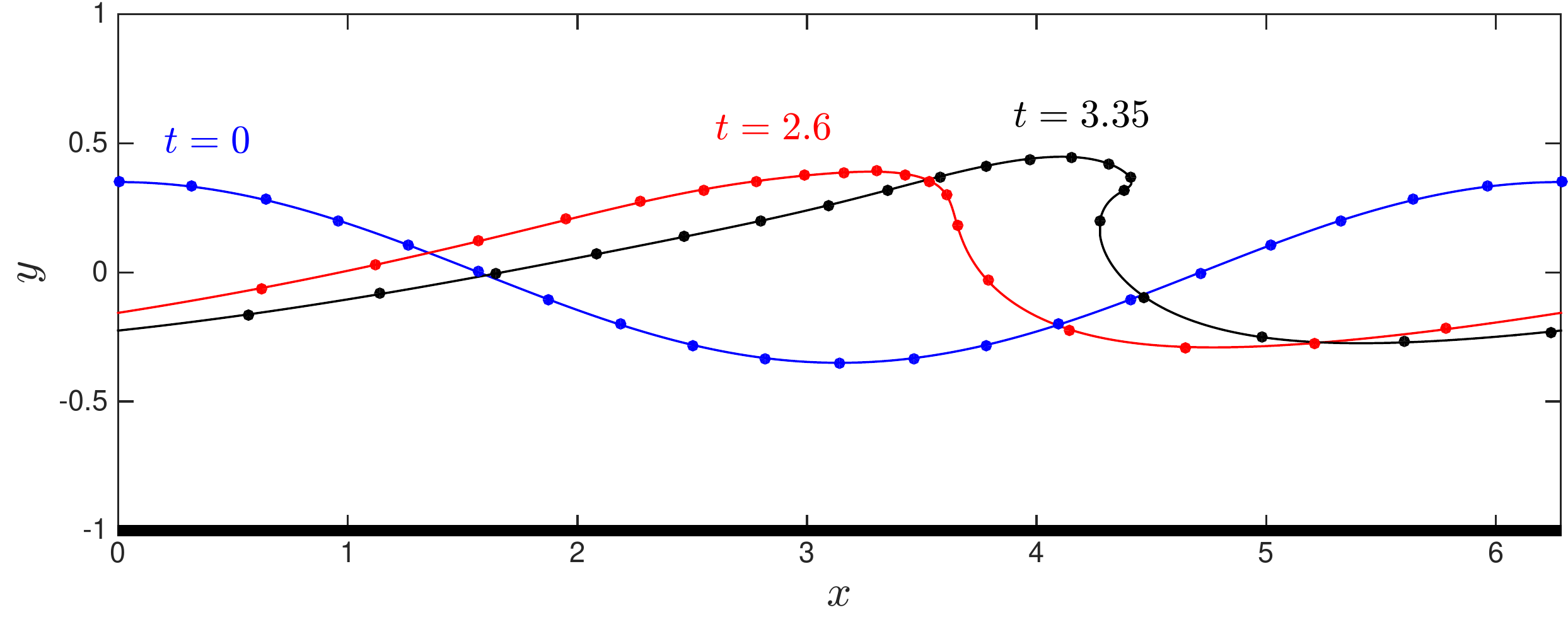}
\caption{Profile of a breaking wave over a flat bottom $y = -1$ at three different times: $t = 0$, $2.6$ and $3.35$. Dot markers correspond to material points, which are distributed at equal distances at initial time; for a better visualization we display only a few markers. The free surface is strongly compressed at the overhanging tip, as seen by the increasing marker density, while it gets stretched at the front slope on the right.}
\label{fig1}
\end{figure}

\begin{figure}
\centering
\includegraphics[width=0.4\textwidth]{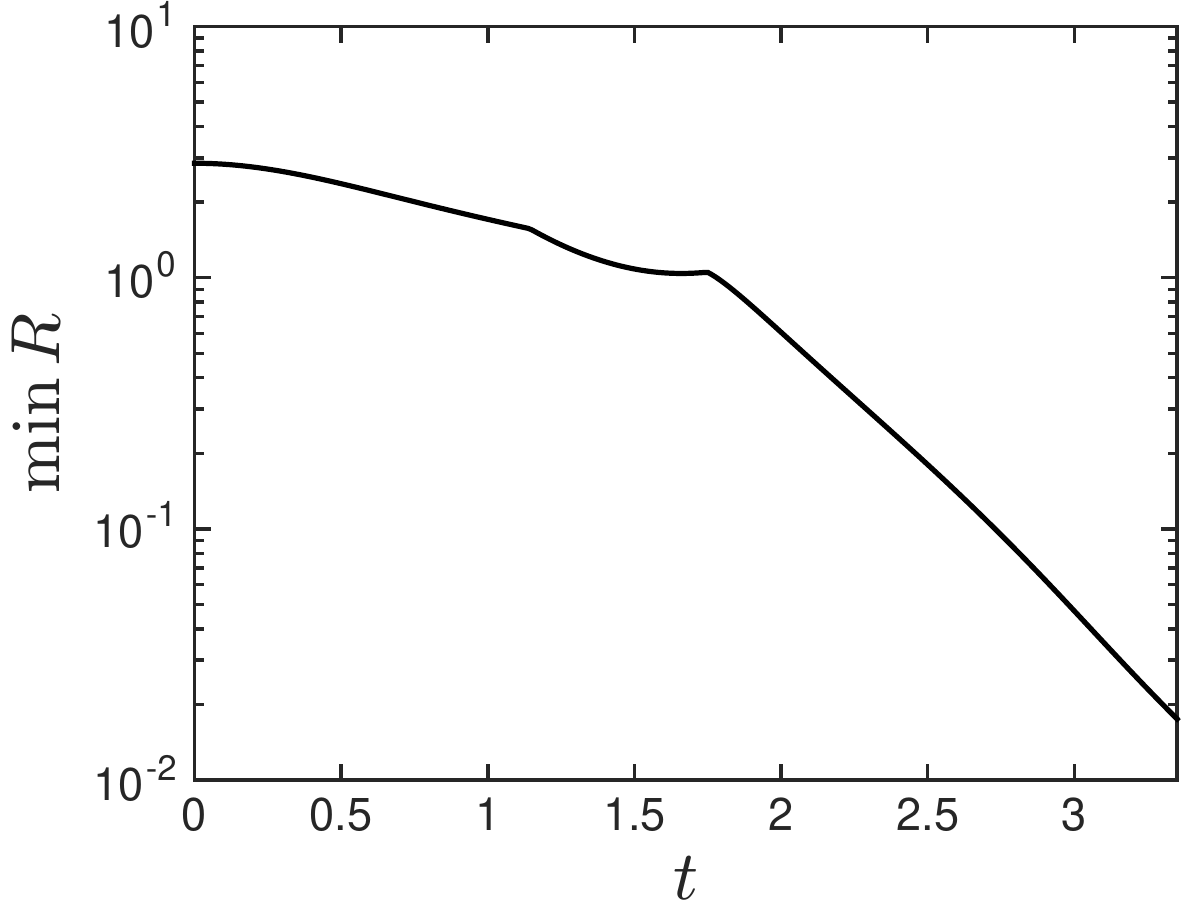}
\caption{Temporal dependence of the minimum curvature radius $R$ along the free surface; logarithmic vertical scale.}
\label{fig9}
\end{figure}

\begin{figure}
\centering
\includegraphics[width=0.45\textwidth]{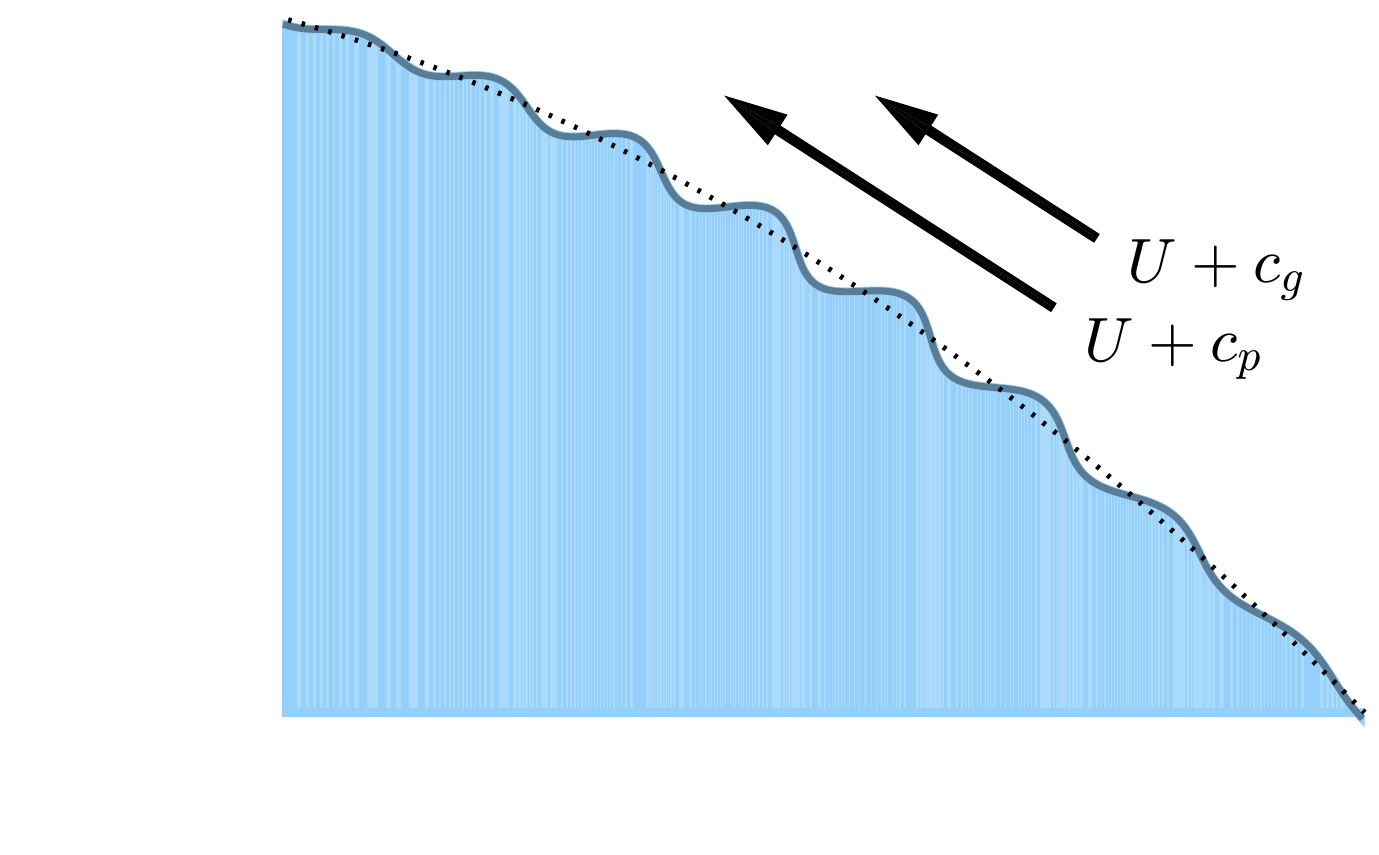}
\vspace{-5mm}
\caption{Small ripple (solid blue line) traveling on top of the unperturbed wave profile (dotted line). Motion of the ripple is approximately described by the sum of the unperturbed flow speed $U$ and the relative group/phase speeds.}
\label{fig10}
\end{figure}

The curvature of the profile 
displayed in Fig.~\ref{fig1} increases rapidly at the overhanging tip. The plot of its minimal curvature radius 
as a function of time is presented in 
a logarithmic vertical scale in Fig.~\ref{fig9},
demonstrating the nearly exponential decrease at later times. Similar solutions were reported in many earlier studies, as for example in  \citep{baker1982generalized,peregrine1983breaking,grilli1990corner,baker2011singularities}.  There 
exist initial conditions that can be rigorously tracked until a splash singularity (e.g. intersection of the wave tip with the bottom) 
occurs in finite time~\citep{castro2012splash}. But this strong overturning is not the main goal of our work.
Considering the 
incipient breaking wave as 
 the large-scale underlying flow, our focus here will be on the study of much shorter
small-amplitude ripples evolving on 
the surface of such  steepening wave profiles (see Fig.~\ref{fig10}).

Small-amplitude ripples 
are described, as a first approximation, with equations linearized about the time-dependent unperturbed wave solution. Let us introduce a local (arc length) coordinate $s$ along the surface, which defines the surface spatial coordinates as $x(s,t)$ and $y(s,t)$. Then it is convenient to consider ripple perturbations in the form of a slowly modulated wavetrain \citep{bretherton1968wavetrains,peregrine1976interaction}
with frequency $\omega(s,t)$, a carrier wavenumber $k(s,t)$ and amplitude $a(s,t)$. 
The regime of interest is such that these ripple parameters may vary with position and time,
where appreciable changes are observed after many periods ($2\pi/\omega$) and many wavelengths $(2\pi/k)$. Then, in the first approximation, the underlying flow due to the wave steepening is
locally constant with respect to the ripple, but accelerating in time. 
The frequency and wavenumber of the ripple can be derived from the phase function $\theta(s,t)$ by
	\begin{equation}
	\omega = -\frac{\partial\theta}{\partial t},\quad
	k = \frac{\partial\theta}{\partial s}.
	\label{eqN2}
	\end{equation}
If $U(s,t)$ is the velocity at the fluid surface with respect to coordinate $s$, then 
as mentioned in the introduction, one has~\citep{bretherton1968wavetrains}
	\begin{equation}
	\omega = Uk+\Omega,
	\label{eqN2b}
	\end{equation}
where $\Omega(s,t)$ is the local intrinsic frequency of the modulated Fourier mode in the Lagrangian reference frame.

\begin{figure}
\centering
\includegraphics[width=0.8\textwidth]{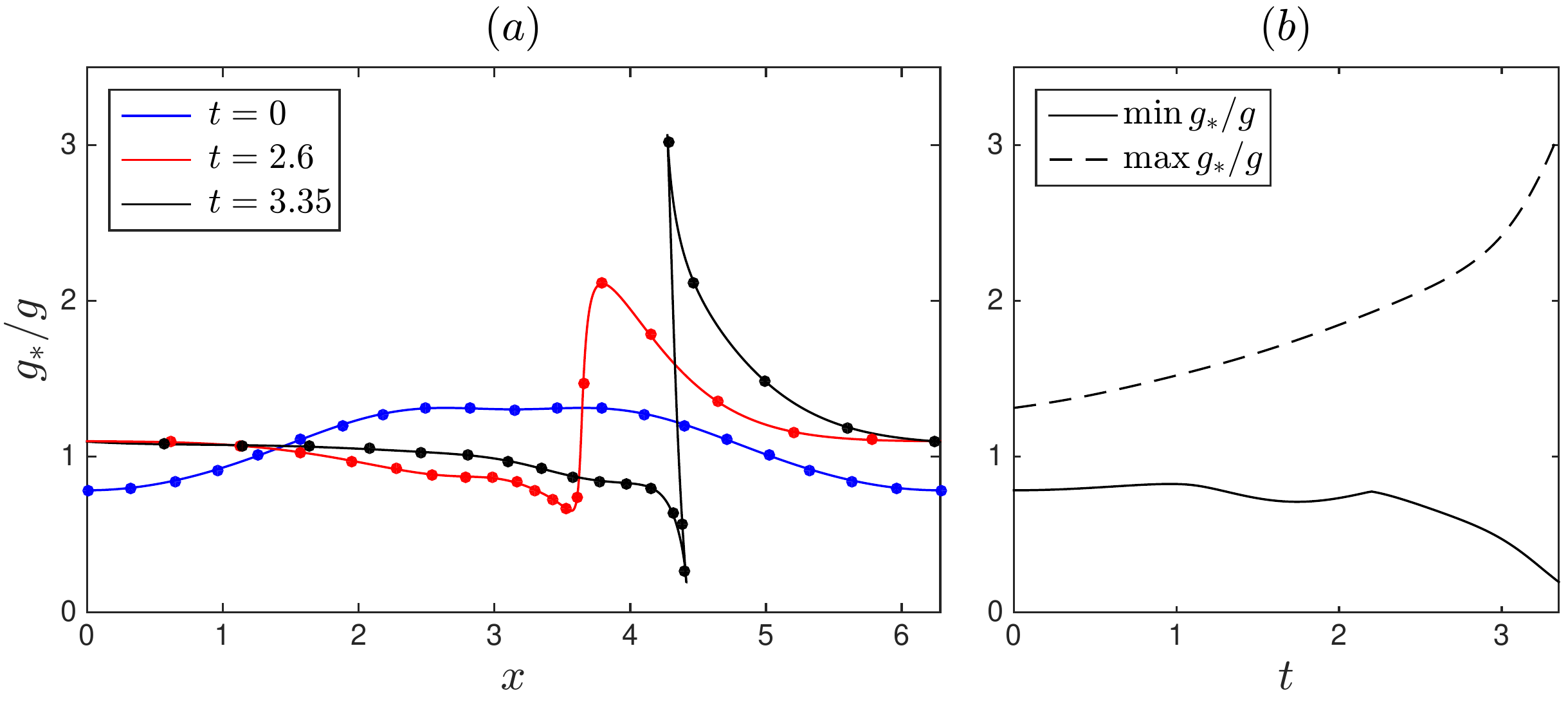}
\caption{(a) Profiles of the effective gravity $g_*$ (in units of $g$) along the water surface at different times; times and selected markers are the same as in Fig.~\ref{fig1}. (b) Minimum and maximum of the effective gravity at the water surface as functions of time. }
\label{fig8}
\end{figure}

Since the deep-water approximation can be used for short wavelengths, the ripple's  dispersion relation in a 
first approximation  is given  by \citep[\S12]{landau1987fluid}
	\begin{equation}
	\Omega = \sqrt{g_* k}.
	\label{eq4}
	\end{equation}
Here $g_*(s,t)$ is the effective (intrinsic) gravity acting on the ripple in the local reference frame,
as its background flow is moving towards overturning.
The effective gravity is defined as $\mathbf{g}_* = \mathbf{g}-\mathbf{a}$, where $\mathbf{a} = (D\mathbf{v}/Dt) = -(\nabla p)/\rho+\mathbf{g}$ is the material acceleration at the surface, under the Euler equations (\ref{eq1}). As a result, we have 
	\begin{equation}
	\mathbf{g}_* = \frac{\nabla p}{\rho} = -g_*\mathbf{n},
	\label{eq3b}
	\end{equation}
where $\mathbf{n}$ is the surface normal vector. Recall that the pressure is constant at the free boundary and, thus, its gradient is orthogonal to the surface. For the numerical results in Fig.~\ref{fig1}, the value of $g_*$ along the wave profile at different times is shown in Fig.~\ref{fig8}(a) 
with the time dependence of its minimum and maximum presented in Fig.~\ref{fig8}(b). At the final time, $g_*$ varies from the minimum $g_* \approx 0.2g$ at the wave tip to the maximum $g_* \approx 3g$ at the foot 
of the wave.  The value of $g_*$ remains positive at all times in our simulation. As proved by~\cite{wu1997well}, the positivity of $g_*$ holds as long as the interface is non-selfintersecting. 
Therefore we are in the stable Rayleigh-Taylor regime.
The local phase speed $c_p$ and group speed $c_g$ are defined as
	\begin{equation}
	c_p  = \frac{\Omega}{k} = \sqrt{\frac{g_*}{k}},\quad
	c_g = \frac{\partial\Omega}{\partial k} = \frac{1}{2}\sqrt{\frac{g_*}{k}},
	\label{eq4c}
	\end{equation}
in the Lagrangian reference frame.

\section{Wave action and the adiabatic approximation for ripple evolution}\label{sec3}

The consistency conditions for second derivatives of phase in (\ref{eqN2}) yield the relation~\citep{bretherton1968wavetrains}
	\begin{equation}
	\frac{\partial k}{\partial t}+\frac{\partial\omega}{\partial s} = 0,
	\label{eqN4}
	\end{equation}
which can be written using (\ref{eqN2b}) and (\ref{eq4c}) as the conservation law
	\begin{equation}
	\frac{\partial k}{\partial t}+\frac{\partial}{\partial s}\left[(U+c_p)k\right] = 0.
	\label{eqN4b}
	\end{equation}
In the adiabatic approximation, i.e., when the temporal and spatial scales of the ripple are much smaller than the scales of the underlying flow (here the unperturbed wave), one also has the conservation law~\citep{bretherton1968wavetrains}
	\begin{equation}
	\frac{\partial}{\partial t}\frac{E}{\Omega}+\frac{\partial}{\partial s}\left[(U+c_g)\frac{E}{\Omega}\right] = 0,
	\label{eqN4c}
	\end{equation}
for the wave action density $E/\Omega$. 
Here $E$ is the ripple energy, which can be obtained by considering a linear wave of amplitude $a$ with the effective gravitational acceleration $g_*$.  
Considering the time-averaged values over an oscillation period and equipartition of the kinetic and potential energies, the local energy density of the ripple is written as~\citep{landau1987fluid}
	\begin{equation} 
	E = \frac{1}{2}\rho g_*a^2.
	\label{eqN4d0}
	\end{equation}
Note that the energy of the entire system is conserved, which means that the adiabatic changes just described reflect the energy exchange between the large-scale motion of the overturning wave and small-scale ripples on the water surface.

The two conservation laws (\ref{eqN4b}) and (\ref{eqN4c}) with the expressions (\ref{eq4}), (\ref{eq4c}) and (\ref{eqN4d0}) define the evolution of the local wavenumber $k$ and amplitude $a$ of the ripple. We will now show that these equations can be solved approximately for the small ripples, when the ripple wavelength $\ell = 2\pi/k$ is considered 
small compared to the scale of the unperturbed nonlinear wave. As it follows from (\ref{eq4c}), such an assumption implies that the ripple phase and group velocities are small compared to the local flow speed,
	\begin{equation}
	c_p \ll U,\quad c_g \ll U.
	\label{eqN4d}
	\end{equation}
Hence, equations (\ref{eqN4b}) and (\ref{eqN4c}) in a first approximation reduce to the form
	\begin{equation}
	\frac{\partial k}{\partial t}+\frac{\partial}{\partial s}(Uk) = 0,\quad
	\frac{\partial}{\partial t}\left(\frac{E}{\Omega}\right)
	+\frac{\partial}{\partial s}\left(U\,\frac{E}{\Omega}\right) = 0.
	\label{eqN4f}
	\end{equation}
Recall that $U(s,t)$ in these equations is the local flow speed on the surface of unperturbed steepening wave; for the linearized formulation, it is not affected by a small-amplitude ripple motion.

Both equations in (\ref{eqN4f}) have the form of the continuity equation 
	\begin{equation}
	\frac{\partial \sigma}{\partial t}+\frac{\partial}{\partial s}(U\sigma) = 0.
	\label{eqN4g}
	\end{equation}
Consider 
	\begin{equation}
	t = 0:\quad \sigma(x) \equiv 1 
	\label{eqN4gIC}
	\end{equation}
to represent an initial uniform marker (material tracer) distribution along the free surface; see Fig.~\ref{fig1}. In this case
the 
(marker density) function $\sigma(s,t)$ describes the stretching (for $\sigma < 1$) and compression (for $\sigma > 1$)
of these material markers along the free surface in time. 
By using  (\ref{eqN4f}) and (\ref{eqN4g}), one can check that 
	\begin{equation}
	\frac{D}{Dt}\left(\frac{k}{\sigma}\right) = 0,\quad
	\frac{D}{Dt}\left(\frac{E}{\sigma\Omega}\right) = 0,
	\label{eqN4gg}
	\end{equation}
where the $D/Dt = \partial/\partial t + U\partial/\partial s$ is the material derivative. In other words, the following quantities 
	\begin{equation}
	\frac{k}{\sigma} = const,\quad
	\frac{E}{\sigma\Omega} = const,
	\label{eqN4h}
	\end{equation}
are invariant along Lagrangian trajectories at the fluid surface. These two quantities, which refer to the local ripple properties, represent approximate (adiabatic) Lagrangian invariants on the free surface. Nonlinear effects of compression are now built-in to expressions (\ref{eqN4h}). 

The first relation in (\ref{eqN4h}) implies that the ripple wavelength $\ell = 2\pi/k$ changes proportionally to $1/\sigma$, i.e.,
	\begin{equation}
	\frac{\ell}{\ell_0} = \frac{1}{\sigma},
	\label{eqN4i}
	\end{equation}
where the zero subscript denotes the initial length value at $t = 0$; recall that $\sigma_0 = 1$ due to (\ref{eqN4gIC}). 
This formula  captures the physical feature that the ripple travels along the Lagrangian trajectory, while stretching  
or compressing according to the material marker's dynamics. For interpreting the second relation in (\ref{eqN4h}), recall that $E/\Omega$ was defined as the wave-action density per unit surface length. Therefore, $E/(\sigma\Omega)$ is the conserved Lagrangian wave-action density, corresponding to the unit surface length at the initial time.

The conserved wave action in (\ref{eqN4h}) will capture the explosive instability. 
It can be written using the dispersion relation (\ref{eq4}) for the intrinsic frequency and the energy density expression (\ref{eqN4d0}) as
	\begin{equation}
	\frac{E}{\sigma\Omega} = 
	 \frac{\rho g_*a^2}{2\sigma\sqrt{g_* k}} 
	 = \frac{\rho \ell_0^{1/2}}{2^{3/2}\pi^{1/2}}\frac{g_*^{1/2}a^2}{\sigma^{3/2}},
	\label{eqN4h2}
	\end{equation}
where in the last equality we used (\ref{eqN4i}) to express  $k = 2\pi/\ell = 2\pi\sigma/\ell_0$.
Since the first factor in the last expression of (\ref{eqN4h2}) is constant, the conservation property (\ref{eqN4h}) yields
	\begin{equation}
	\frac{g_*^{1/2}a^2}{\sigma^{3/2}} = const. 
	\label{eqN4h2exA}
	\end{equation}
Evaluating the constant from the initial time, one obtains
	\begin{equation}
	\frac{a}{a_0} = \left(\frac{\sigma^3 g_{*0}}{g_*}\right)^{1/4},
	\label{eqN4k}
	\end{equation}
where $a_0$ and $g_{*0}$ are, respectively, the values of $a$ and $g_*$ at $t = 0$. 
Combining expressions (\ref{eqN4i}) and (\ref{eqN4k}), we express the ripple steepness (the ratio of height to wavelength) as
	\begin{equation}
	S = \frac{2a}{\ell} = \frac{2\sigma a}{\ell_0}
	\label{eq10RS}
	\end{equation}
with our final formula
	\begin{equation}
	\frac{S}{S_0} = \left(\frac{\sigma^7 g_{*0}}{g_*}\right)^{1/4},
	\label{eq10}
	\end{equation}
where $S_0 = 2a_0/\ell_0$.

Recall that all relations (\ref{eqN4h})--(\ref{eq10}) are deduced along Lagrangian trajectories at the water surface. 
As will be shown numerically, it is remarkable that these
rather simple formulas encompass the full action of the changing large-scale wave profile on 
the passive small-scale ripple with several nontrivial implications. First, the ripple steepness is fully controlled by the surface compression ratio $\sigma$ and the local effective gravity $g_*$. 
Due to the rather large exponent of the term $\sigma^{7/4}$ in (\ref{eq10}), the marker density function has a strong  effect.
Going back to Section 2, note that the compression ratio strongly varies within a wave during the breaking process; see, e.g., two  Lagrangian points very close to the wave tip in Fig.~\ref{fig1}.
Second, expression (\ref{eq10}) does not depend on the ripple wavelength, predicting that all ripples steepen 
at the same rate. In particular, this justifies the use of formula (\ref{eq10}) for a ripple in the form of a general short wave-length 
modulated perturbation on top of the original wave profile.

\section{Numerical results}\label{sec5}

\begin{figure}
\centering
\includegraphics[width=0.8\textwidth]{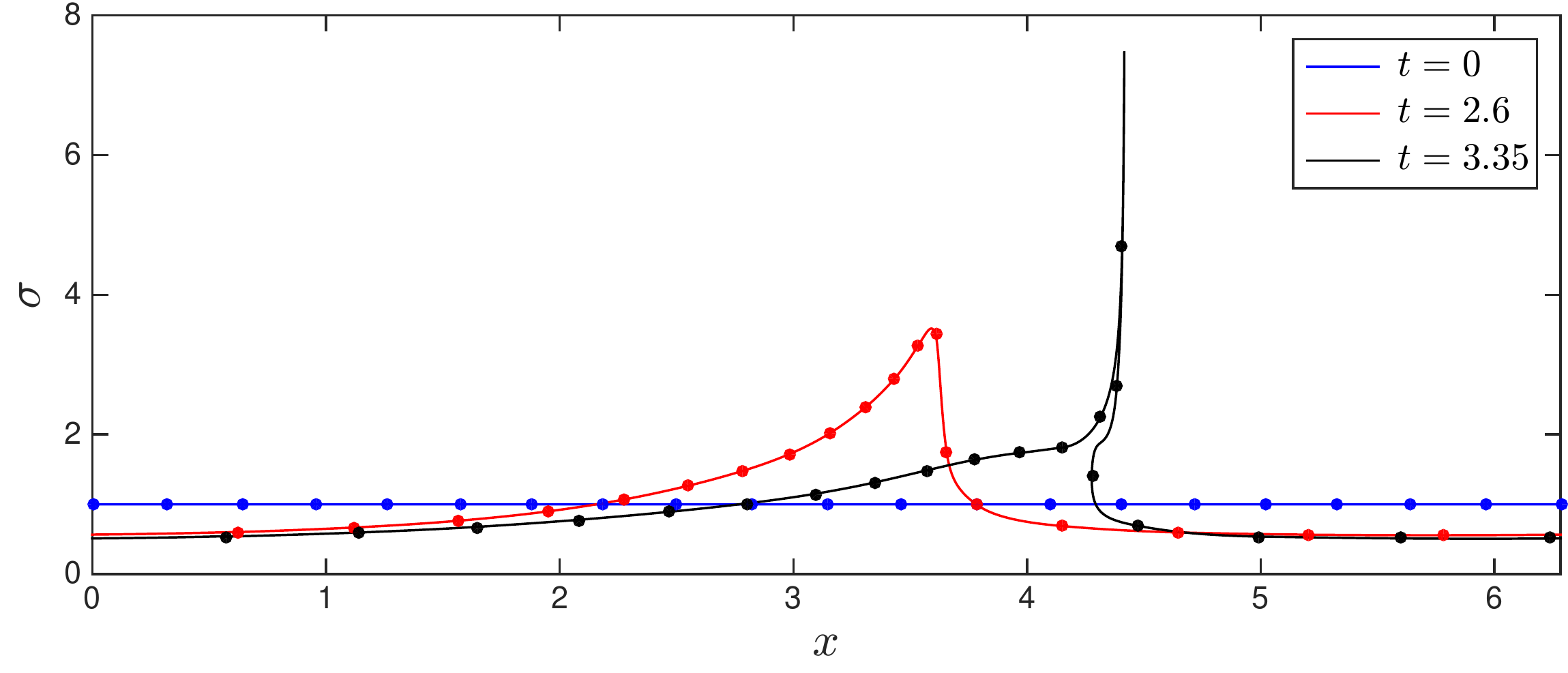}
\caption{Profile of the surface marker density $\sigma$ vs. the corresponding horizontal coordinate $x$ at different times. For $t = 3.35$ the graph is multi-valued due to the overhanging wave profile; see Fig.~\ref{fig1}.}
\label{fig11}
\end{figure}

\begin{figure}
\centering
\includegraphics[width=0.7\textwidth]{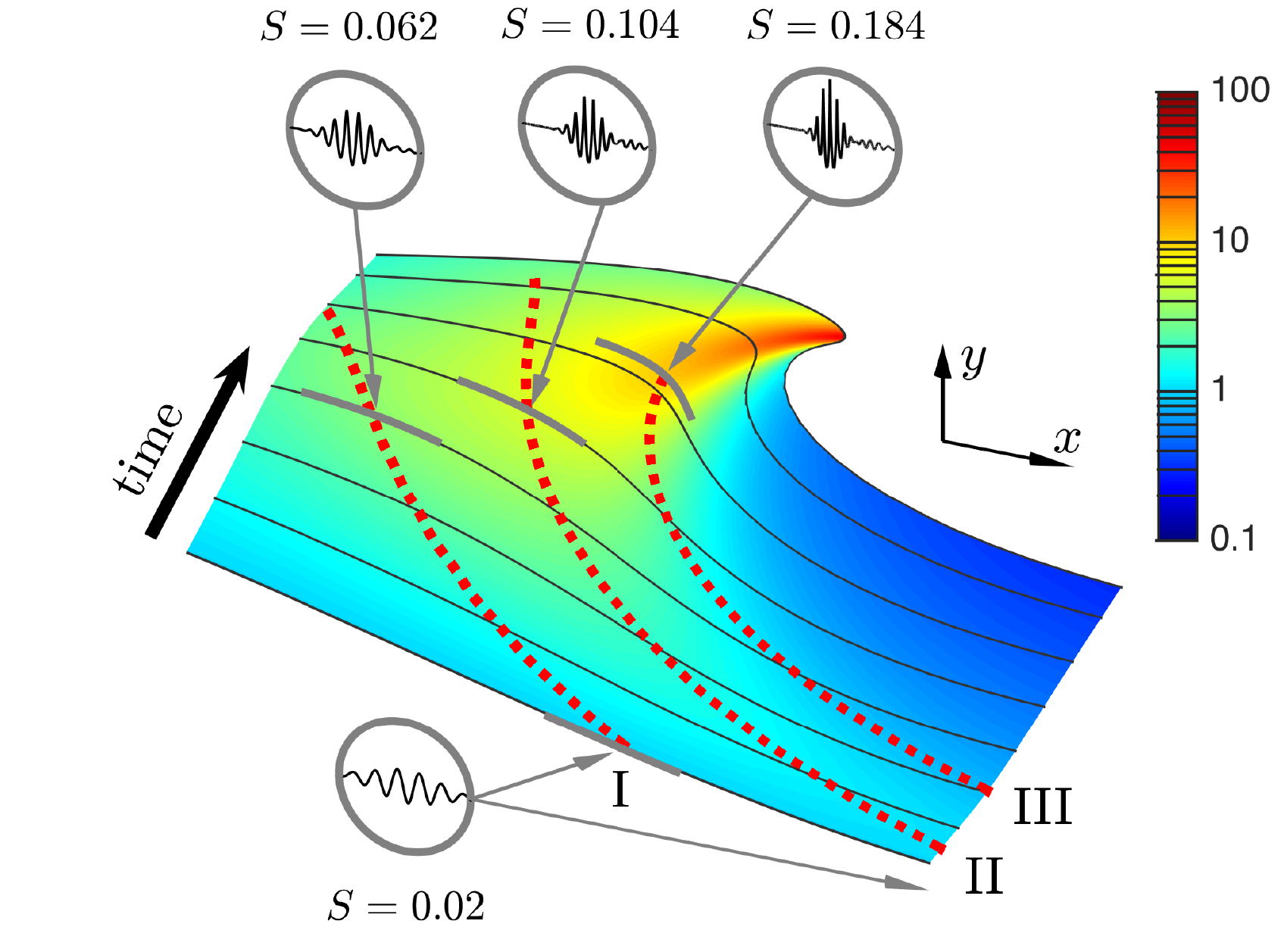}
\caption{Color (in log scale) shows the change of ripple steepness, $S(t)/S_0$, at different points of the wave profile. The steepness decreases at the wave foot (blue) and then increases by almost two orders of magnitude at the wave tip (red). Black curves on a surface show the wave profiles at times $t = 0,\ 0.5,\ 1,\ldots,3$ and the final time $3.35$. Dashed red curves indicate the trajectories of small ripples (centers of Gaussian wave packets) located initially at (I) $x = 2$, (II) $x = 2.9$ and (III) $x = 3.8$; the simulation in case III was stopped at $t = 2.575$. The insets present the shape and steepness of these ripples at initial time (circle at the bottom) and at different later times (circles at the top), shown with the subtracted background profile and magnified vertical scale 100:1.}
\label{fig4}
\end{figure}

\begin{figure}
\centering
\includegraphics[width=0.72\textwidth]{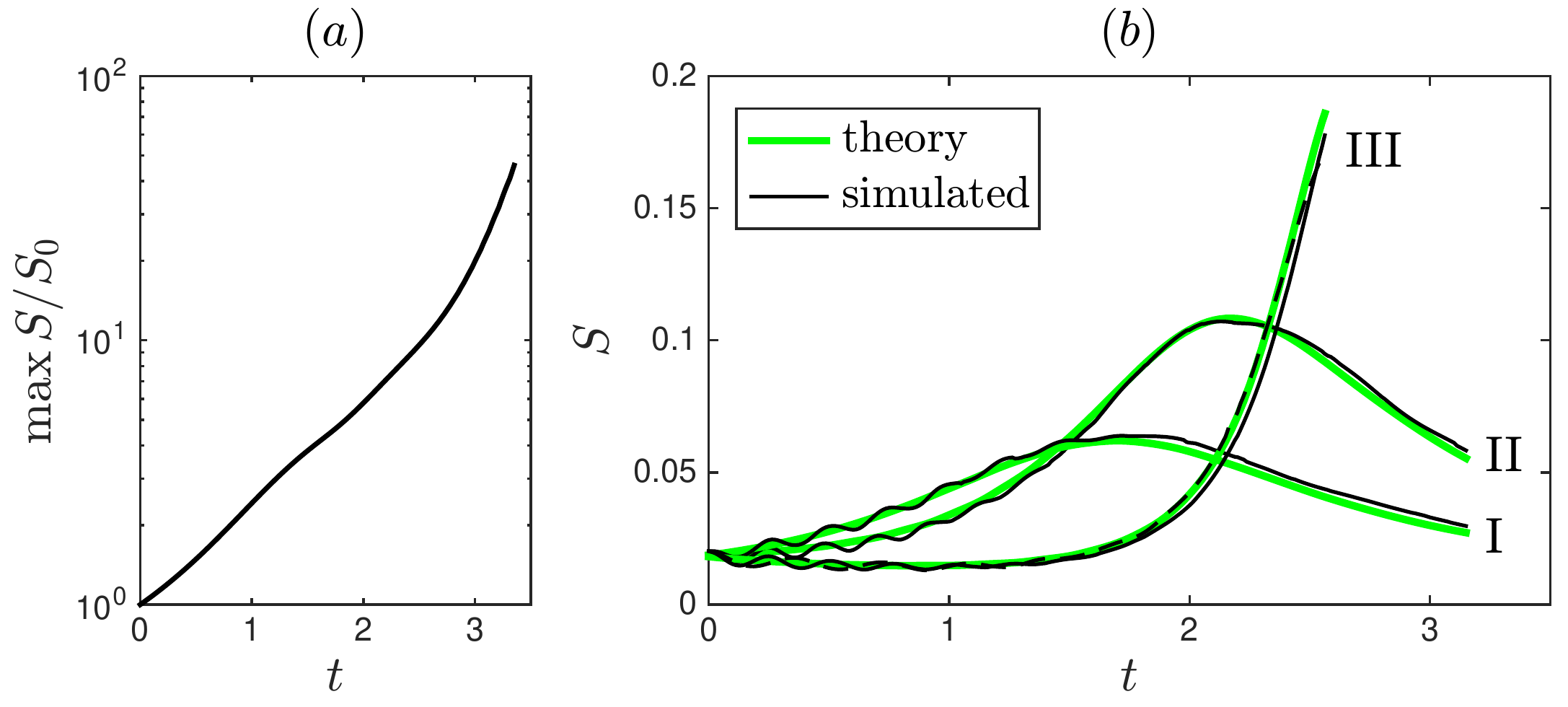}
\caption{(a) Increase of ripple steepness (maximum value within a wave), demonstrating a super-exponential growth at later times; vertical log-scale. The ripple steepness increases about 50 times. (b) Change of steepness with time for three different Gaussian ripples, see also Fig.~\ref{fig4}. Adiabatic theoretical prediction is compared with numerical simulations. The dashed line corresponds to the simulation of a ripple with a twice larger wavelength.}
\label{fig5}
\end{figure}

In simulations, we consider the ripples in the form of short Gaussian wave packets. 
According to relations (\ref{eqN4d}), such wave packets follow approximately the Lagrangian fluid trajectories at the surface.
Expression (\ref{eq10}) for the ripple steepness depends only on the unperturbed solution in Fig.~\ref{fig1} and, thus, can be evaluated at every point of the wave profile. The profiles of the effective gravity $g_*$ were already shown in Fig.~\ref{fig8}. 
Numerically computed   profiles 
of the surface marker density  $\sigma$ are presented in Fig.~\ref{fig11}, 
demonstrating the
strong  compression (large $\sigma$) at the wave tip, and some stretching ($\sigma < 1$) at the wave foot. 
Together, these results provide the change in ripple steepness along the wave profile, which
is depicted in a logarithmic color-scale in Fig.~\ref{fig4}. 
Here one observes an explosive growth of ripple steepness by almost 50 times near the overhanging wave tip (red color). At the foot of the wave the steepness decreases (blue color) depleting the surface roughness. Note that our results are obtained within the two-dimensional model. The third dimension is not crucial for the dynamics of short-wave ripples, since their speeds in the direction transversal to the wave can be neglected due to relations (\ref{eqN4d}). The steepness increases super-exponentially in time as shown in Fig.~\ref{fig5}(a). 
Such explosive behavior distinguishes the 
present adiabatic steepening mechanism
from common instabilities featuring an exponential growth. 

For verification of our theory, we added small ripples in the form of Gaussian wave packets with steepness $S \approx 0.02$ and wavenumber $k = 128$ located at different parts of the initial 
nonlinear wave profile and repeated our numerical simulations; see the Appendix \S\ref{secA3} for numerical aspects of creating such perturbed initial conditions. Trajectories of these packets, computed as centers of their envelopes, are shown in Fig.~\ref{fig4} by red dotted lines. Their shapes at different times are displayed in the insets, where the background profile was subtracted and the vertical scale was magnified with the ratio 100:1. One can see that the ripple steepness increases as a combination of two factors: the decrease of ripple wavelength and the increase of its amplitude. Recall that such ripples are small perturbations, which do not affect considerably the overturning wave profile.

Fig.~\ref{fig5}(b) demonstrates an excellent agreement between the numerical steepness measured at the center of each packet and the theoretical (adiabatic) prediction (\ref{eq10}). In case III of Fig.~\ref{fig5}(b), we observe both the decrease of steepness (depleting the surface roughness) at early times followed by its sudden increase by more than one order of magnitude. We also performed the simulation for the Gaussian packet with a twice larger wavelength in the case III; see the dashed black line in Fig.~\ref{fig5}(b). This confirmed the independence on wavelength for the ripple evolution predicted theoretically by Eq.~(\ref{eq10}). 

It is instructive to provide the reader with an example of the corresponding dimensional variables for our simulations: one can take the water depth to be $5$m,  the wave height reaching $3$m and the initial ripples 
having a wavelength of $0.25$m. Such values are typical for ocean waves and, by their orders of magnitude, one can expect that surface tension does not play a significant role at the initial stage of ripple steepening. 

\begin{figure}
\centering
\includegraphics[width=0.65\textwidth]{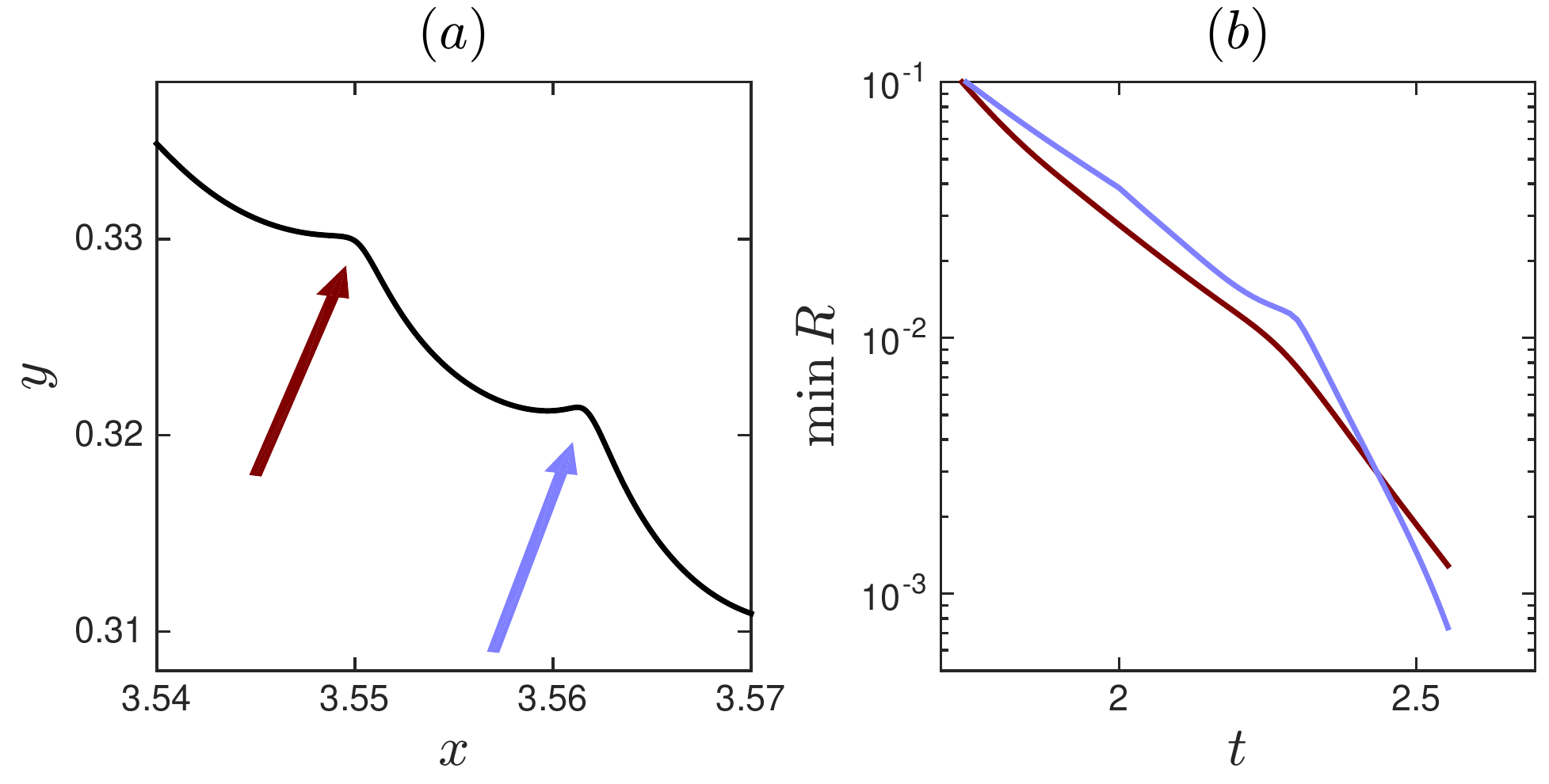}
\caption{(a) Onset of angle formation at ripple crests. Shown is a segment of the wave profile at late time $t = 2.575$ in case III; see Fig.~\ref{fig4}. (b) Time dependence of curvature radius at ripple crests indicated in the left panel; vertical log-scale.}
\label{fig6}
\end{figure}

Due to the rapidly increasing steepness, small ripples can reach a strongly nonlinear regime, which 
should typically happen in the yellow-to-red region around the wave tip in Fig.~\ref{fig4}. Considering as an example the case III, we 
observe in Fig.~\ref{fig6}(a) that singularities (sharp angles) are about to form at the ripple crests when the steepness gets close to $S \approx 0.18$. At these points, the curvature radius is decreasing at least exponentially with time; 
see Fig.~\ref{fig6}(b). Therefore surface tension becomes important when the curvature gets large. 
In our simulations, according to Fig.~\ref{fig6}, this can be expected around the time $t \approx 2.6$, i.e., prior to overturning in Fig.~\ref{fig1}.

\begin{figure}
\centering
\includegraphics[width=0.95\textwidth]{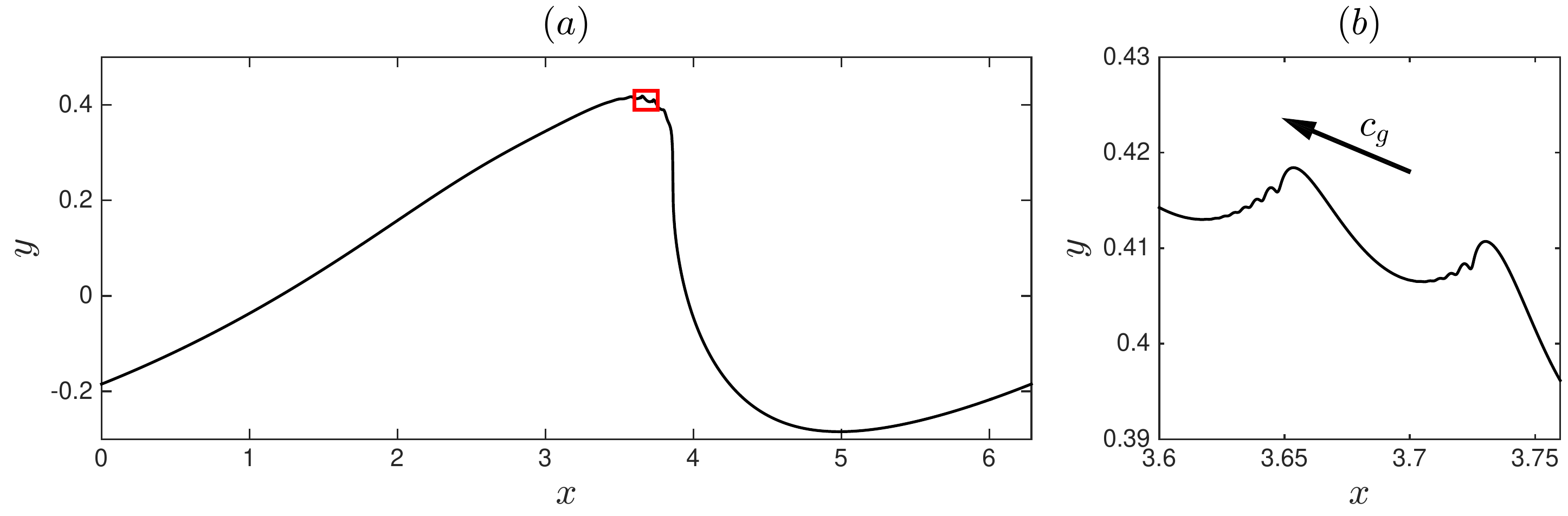}
\caption{Ripple shape at late time $t = 2.85$ for the simulation with the dimensionless surface tension coefficient $\gamma = 6\times 10^{-6}$. The initial wave packet corresponds to the case III, where we increased four times the ripple wavelength to improve numerical resolution. Figure (b) represents the magnified region of the ripple shown with the small red square on left panel. The arrow indicates the direction of the (small) group speed of the ripple relative to the fluid. One can see the formation of parasitic capillary waves developing in front of the ripple crests.}
\label{fig7}
\end{figure}

In Fig.~\ref{fig7} we present numerical results for our model, now taking surface tension into account. In this case, the dynamical (stress balance) condition for the pressure at free-surface yields
	\begin{equation}
	p = P_{atm}+\gamma/R,
	\label{eq_ST}
	\end{equation}
where $\gamma$ is the surface tension coefficient and $R$ is the curvature radius of the free surface. The 
simulation 
producing Fig.~\ref{fig7} was performed with the dimensionless value $\gamma = 6\times 10^{-6}$, which corresponds to a realistic value of surface tension for breaking waves of moderate height; see the Appendix \S\ref{secA1} for details of the numerical method.
	
The resulting magnified profile in Fig.~\ref{fig7}(b) reveals the secondary ``ripple breaking" when the curvature at the ripple crests becomes large, followed by the generation of the so-called ``parasitic'' capillary waves~\citep{ceniceros1999dynamic}. Such capillary waves are known to form near the crests of nonlinear gravity waves in a resonant manner~\citep{longuet1995parasitic}, and could be a mechanism of whitecapping~\citep{dyachenko2016whitecapping}. 
 
\section{Conclusions}

We developed the asymptotic theory describing the coupling of large-scale wave breakers to small-scale surface ripples traveling on its surface. This theory is constructed using the analogy with wavetrains propagating on a free surface of water that are influenced by large-scale currents. However, in contrast to the latter case where the wavetrain behavior is governed by the intrinsic frequency through the approximate conservation of the wave action~\citep{bretherton1968wavetrains,peregrine1976interaction}, in our case two distinct quantities are important: the intrinsic frequency and the nonlinear intrinsic gravity. Both these quantities are introduced in the local Lagrangian reference frame at each point of the free surface. They define the wave action as an adiabatic Lagrangian invariant for the potential ideal flow. 

Unlike the common hydrodynamic instabilities, this mechanism predicts super-exponential growth of ripple steepness in time, resulting from simultaneous decrease of wavelength and increase of amplitude, with excellent quantitative agreement between the developed theory and numerical simulations. 
When taking capillary effects into account, our simulations anticipate the small-scale ``ripple breaking" along the water surface revealing the increasing complexity of the subsequent nonlinear process. 
The proposed theory is asymptotic and requires further development both for its rigorous justification (along with the underlying adiabatic approach for gravity waves) and for a better understanding, for example, of the parameter range where the instability takes place. The study of ripples in nonlinear regime is also of interest, e.g. from the perspective of finite-time singularities~\citep{longuet1983bubbles,zeff2000singularity}.  

\begin{figure}
\centering
\includegraphics[width=0.8\textwidth]{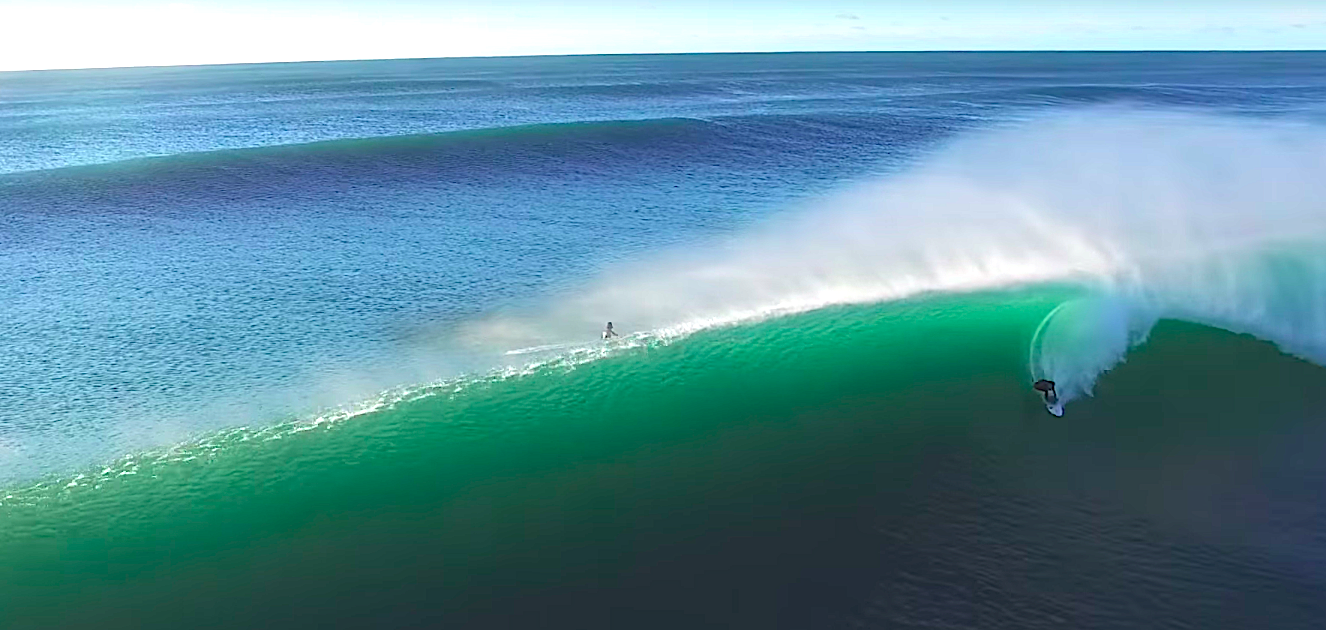}
\caption{Breakdown of a smooth water surface in a plunging ocean wave (figure courtesy of Keahi de Aboitiz).}
\label{fig2}
\end{figure}

We expect that our results may contribute to deeper understanding of multi-scale aspects of wave breaking, such as surface fragmentation and whitecapping. This expectation is based on the observation that the ripple instability we described occurs in the ideal Euler setting prior to the onset of parasitic (capillarity) oscillations~\citep{longuet1995parasitic}; the latter are conjectured to be a mechanism of bubbles formation~\citep{dyachenko2016whitecapping}. Figure~\ref{fig2} shows the formation of small white sprays (left side) developing later into a strongly fragmented wave tip (right side); a closer look reveals that the appearance of white regions has correlations with the locations of ripples crests.  

\section{Appendix: Numerical method}

In this section we describe briefly the numerical method and provide final equations to be simulated. For more details see~\citep{dyachenko1996analytical,ribeiro2017flow} and also \citep{zakharov2002new} for an alternative way to represent the same equations of motion. 

Having three dimensional parameters, the channel period $L$ [m], the acceleration of gravity $g$ [m/s$^2$] and the fluid density $\rho$ [kg/m$^3$], we define the dimensionless variables for space, time, velocity and pressure as
	\begin{equation}
	x \mapsto x\,\frac{2\pi}{L}, \quad
	y \mapsto y\,\frac{2\pi}{L}, \quad
	t \mapsto t\sqrt{\frac{2\pi g}{L}}, \quad
	\mathbf{v} \mapsto \mathbf{v}\sqrt{\frac{2\pi}{gL}}, \quad
	p \mapsto p\,\frac{2\pi}{\rho gL}.
	\label{eq1.0d}
	\end{equation}
In the new variables, the dimensional parameters become $L \mapsto 2\pi$, $\rho \mapsto 1$ and $g \mapsto 1$. For the surface tension coefficient, this procedure yields the dimensionless parameter $\gamma \mapsto 4\pi^2\gamma/(\rho gL^2)$.  
In dimensionless variables, we consider a potential flow with period $2\pi$ in the horizontal direction $x$, over a flat rigid bottom at $y = -1$.

\subsection{Basic equations}
\label{secA1}

Using the methods of complex analysis, it is possible to write Euler equations in terms of three real scalar functions $K(t)$, $\hat{A}(\xi,t)$ and $\hat{\varphi}(\xi,t)$, where the dependence on $\xi$ is $2\pi$-periodic; we use the hats to distinguish the functions of $\xi$. These equations have the form \citep{dyachenko1996analytical,zakharov2002new,ribeiro2017flow}
	\begin{equation}
	 K_t = -\frac{1}{2\pi}\int_0^{2\pi} 
	 \frac{\mathsf{R}\hat{\varphi}_\xi}{|\hat{z}_\xi|^2}\,d\xi,
	\label{eq5.1b}
	\end{equation}
	\begin{equation}
	\hat{A}_t 
	= \left[(\mathsf{R}\hat{A}_\xi)-\left(1+\hat{A}_\xi\right)\mathsf{T}\right]
	\frac{\mathsf{R}\hat{\varphi}_\xi}{|\hat{z}_\xi|^2},
	\label{eq5.1}
	\end{equation}
	\begin{equation}
	\hat{\varphi}_t = 
	-\hat{\varphi}_\xi\mathsf{T}\frac{\mathsf{R}\hat{\varphi}_\xi}{|\hat{z}_\xi|^2}
	-\frac{|\hat{\varphi}_\xi|^2-|\mathsf{R}\hat{\varphi}_\xi|^2}{2|\hat{z}_\xi|^2}
	-g\hat{y},
	\label{eq5.2}
	\end{equation}	
where 
	\begin{equation}
	|\hat{z}_\xi| = |\hat{x}_\xi+i\hat{y}_\xi| = \left|1+(1+i\mathsf{R})A_\xi\right|,
	\label{eq2.1cBB}
	\end{equation}
with
	\begin{equation}
	\hat{x}(\xi,t) = \xi+\hat{A}(\xi,t),\qquad \hat{y}(\xi,t) = K(t)-1+\mathsf{R}\hat{A}(\xi,t).
	\label{eq2.1cBB}
	\end{equation}
The operators $\mathsf{R}$ and $\mathsf{T}$ are defined as
	\begin{equation}
	\mathsf{R} \hat f(\xi) = \sum_{m\in\mathbb{Z}}i\tanh ( K m)f_me^{im\xi},\quad
	\mathsf{T} \hat f(\xi) = -\sum_{m \ne 0}i\coth ( K m)f_me^{im\xi},
	\label{eq5.3}
	\end{equation}
for any periodic function $\hat f(\xi) = \sum f_m e^{im\xi}$. 
Here, the shape of the free surface is obtained implicitly as $x = \hat{x}(\xi,t)$ and $y = \hat{y}(\xi,t)$, where
$\xi$ is the auxiliary variable parametrizing the surface.

The complex potential $\Phi(z,t)$ in the fluid domain is given implicitly by
	\begin{equation}
	\Phi = (\hat\zeta+iK)P
	+\mathsf{S}\hat{\varphi},
	\qquad
	z = i(K-1)+\hat\zeta
	+\mathsf{S}\hat{A},
	\label{eq2.1cB}
	\end{equation}
with the operator 
	\begin{equation}
	\mathsf{S}\hat{f} = 
	\sum_{m \in \mathbb{Z}}
	\frac{f_m\exp\left[im(\zeta+iK)\right]}{\cosh Km}.
	\label{eq2.3cB}
	\end{equation}
Here $\zeta = \xi+i\eta$ with $-K \le \eta \le 0$, where the free surface corresponds to $\eta = 0$ and the rigid bottom to $\eta = -K$. 
The velocity field can be obtained from the derivatives of the potential using (\ref{eq1pot}), and then the pressure is given by the Bernoulli equation, $\varphi_t+\left(\varphi_x^2+\varphi_y^2\right)/2+p/\rho+gy = const$. All quantities used in this paper can be computed from the velocity and pressure distributions.

When surface tension $\gamma$ is taken into account, one substitutes equation (\ref{eq5.2}) by~\citep{dyachenko1996analytical} 
	\begin{equation}
	\hat{\varphi}_t = 
	-\hat{\varphi}_\xi\mathsf{T}\frac{\mathsf{R}\hat{\varphi}_\xi}{|\hat{z}_\xi|^2}
	-\frac{|\hat{\varphi}_\xi|^2-|\mathsf{R}\hat{\varphi}_\xi|^2}{2|\hat{z}_\xi|^2}
	-g\hat{y}-\frac{\gamma}{R}.
	\label{eq5.2b}
	\end{equation}	
Here the surface tension is represented by the last term, where the radius of curvature is given by the standard relation $R = (\hat{x}_\xi^2+\hat{y}_{\xi}^2)^{3/2}/(\hat{x}_\xi\hat{y}_{\xi\xi}-\hat{y}_\xi\hat{x}_{\xi\xi})$.

\subsection{Initial conditions and numerical scheme}
\label{secA2}

Initial conditions for system (\ref{eq5.1b})--(\ref{eq5.2}) are obtained as follows. Consider the initial wave profile given by the function $y = y_{ini}(x)$; in our simulations we used  $y_{ini}(x) = 0.35\cos x$. Then, initial condition for the function $\hat{A}(\xi)$ is obtained as $\hat A(\xi) = \mathsf{T}\hat y(\xi)$, where $\hat y(\xi)$ is a limiting point of the iterative scheme $\hat{y}_{n+1}(\xi) = y_{ini}\left(\xi+\mathsf{T}\hat y_n(\xi)\right)$ ($n \to \infty$); see \citep{yu2012exact}. Then for the initial function $\hat{\varphi}(\xi)$, we have $\hat{\varphi}(\xi) = \varphi_{ini}(\hat{x}(\xi))$ with $\hat{x}(\xi) = \xi+\mathsf{T}\hat y(\xi)$, where $\varphi_{ini}(x)$ is the initial value of the real potential at free surface. In our simulations, $\varphi_{ini}(x) = ({0.35}/{\sqrt{\tanh 1}}) \sin x$. Finally, the initial (canonical) depth value is $K = 1+\frac{1}{2\pi}\int_0^{2\pi}\hat{y}(\xi)d\xi$.

In numerical simulations, we use the uniform grid $\xi = 2\pi j/N$ with $j = 0,1,\ldots,N-1$ and apply the spectral method for computing derivatives with respect to $\xi$ as well as for the operators $\mathsf{R}$ and $\mathsf{T}$. Integration in time is done using the fourth-order Runge--Kutta method. To suppress numerical instability at large wavenumbers, which we observed in the simulations, we use the 36th-order smoothing with the Fourier filter, $\exp\left(-36(2k/N)^{36})\right)$, at each time step. This filter was suggested and has been proven to be efficient in high-accuracy numerical simulations of the 3D incompressible Euler equations~\citep{hou2009blow}, and it also worked very efficiently in our simulations. Considering different types of Fourier cutoffs, we checked that the numerical results were not affected by the choice of the filter.

\begin{figure}
\centering
\includegraphics[width=0.45\textwidth]{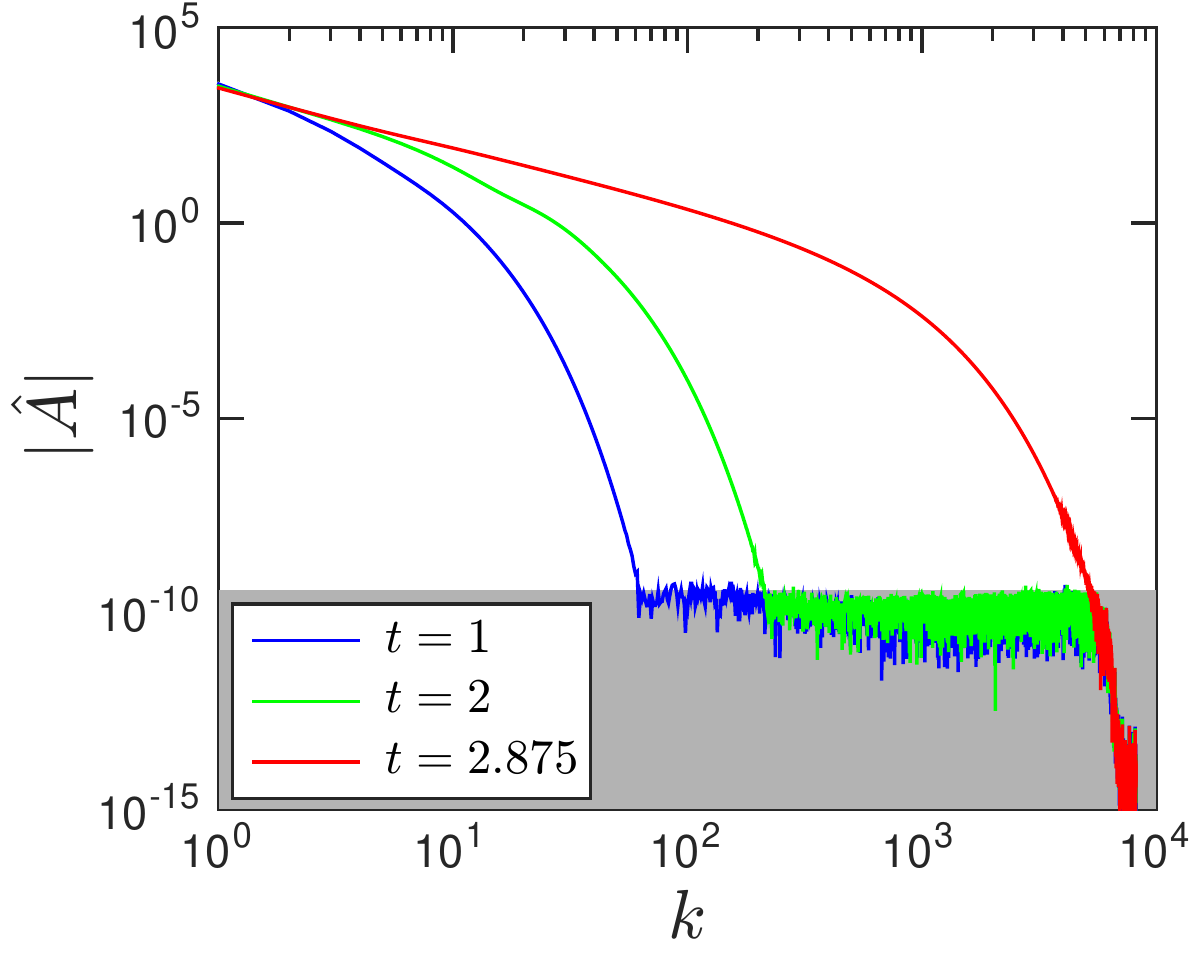}
\caption{Spectrum of numerical wave solution for the Fourier transformed function $\hat{A}(\xi,t)$ at different times. At the final time, the spectrum approaches the limit to the right. Then the grid is refined and the same procedure is repeated.}
\label{figA1}
\end{figure}

We paid special attention to high accuracy of the obtained solution. For this reason we applied the strategy of adaptive mesh refinement similar, e.g., to \citep{agafontsev2015development,dyachenko2016whitecapping} for optimizing the computational performance without affecting the numerical accuracy. We start with $n = 2^{14} = 16384$ grid points and continue the simulation while the solution is only affected by roundoff errors. This can be controlled by checking the wave solution spectrum $\hat{A}$ as shown in Fig.~\ref{figA1}. With the chosen grid the simulation error is dominated by round-off errors until time $t \approx 2.2875$. At this time, we use a very accurate Fourier interpolation to a (twice) larger grid, which extends the spectrum in Fig.~\ref{figA1} to the right. Then the same procedure is repeated with the larger grid, etc. We finished our simulations with the fine grid of $2^{21} = 2097152$ points at time $t = 3.35$. Further increase of the grid is not possible due to computational limitations. Thus, the accuracy of all presented numerical results is kept at the level of round-off errors. We verified that, though simulation can be continued beyond this time with the same grid, the errors increase very fast and the results do not provide any additional insight for our problem.

\subsection{Gaussian wave packets}
\label{secA3}

Using the deep-water linear theory  \citep[\S12]{landau1987fluid}, we have chosen initial conditions for the small-scale ripples as Gaussian wave packets, $y = a_0 e^{-100x^2}\cos kx$ and $\varphi = -a_0 \omega^{-1}e^{-100x^2}\sin kx$. Such packets have the width $0.05$, and we considered the wavenumber $k = 128$ as providing ripples with a small wavelength $2\pi/k \approx 0.049$. The frequency was determined from Eq.~(\ref{eq4}). The amplitude $a_0$ was chosen to give a small ripple steepness $S_0 = 0.02$ at $t = 0$. This wave packet was shifted and superimposed on top of the unperturbed initial wave profile at different locations as specified in the caption of Fig.~\ref{fig4}. In this superposition, a linear interpolation was used to extend the unperturbed velocity potential to the perturbed profile. In some simulations, we also used wider Gaussian packets with $k = 64$.

\vspace{5mm}
\noindent\textbf{Acknowledgements:}
The work of AAM was supported by the CNPq (grant 302351/2015-9). The work of AN was supported by the CNPq (grant 301949/2012-2) and FAPERJ project number E-26/201.164/2014.

\bibliographystyle{jfm}
\bibliography{refs}

\end{document}